\newcommand{\mystyleflag}{0}

\ifnum\mystyleflag=0
\documentclass[12pt]{article}
\pdfoutput=1
\usepackage{jcappub}
\else
\documentclass[journal,article,submit,pdftex,moreauthors]{Definitions/mdpi} 
\firstpage{1} 
\pubvolume{1}
\issuenum{1}
\articlenumber{0}
\pubyear{2024}
\copyrightyear{2024}
\datereceived{ } 
\daterevised{ } % Comment out if no revised date
\dateaccepted{ } 
\datepublished{ } 
\fi

\usepackage{array}
\usepackage{booktabs}
\usepackage{amsmath}
\usepackage{amssymb}
\usepackage{physics}
\usepackage{xcolor}
\usepackage{slashed}
\usepackage{subfigure}
\usepackage{float}
\usepackage{notoccite}
\def\be{\begin{equation}}
\def\ee{\end{equation}}
\def\bea{\begin{eqnarray}}
\def\eea{\end{eqnarray}}

\ifnum\mystyleflag=0
\title{Molecular Learning Dynamics}
\author[1,3]{Yaroslav Gusev}
\emailAdd{yarik.gusev.2000@gmail.com}\author[1,2]{and Vitaly Vanchurin} 
\emailAdd{vitaly.vanchurin@gmail.com}
\affiliation[1]{Artificial Neural Computing, Weston, Florida, 33332, USA}
\affiliation[2]{Duluth Institute for Advanced Study, Duluth, Minnesota, 55804, USA}
\affiliation[3]{Skolkovo Institute of Science and Technology, Moscow, 121205, Russia}
\begin{document}   
\else
% Entropy
\Title{Molecular Learning Dynamics}
\TitleCitation{Molecular Learning Dynamics}
% Authors, for the paper (add full first names)
\Author{Yaroslav Gusev $^{1}$ and Vitaly  Vanchurin $^{1,2}$}
%\longauthorlist{yes}
% MDPI internal command: Authors, for metadata in PDF
\AuthorNames{Yaroslav Gusev and Vitaly  Vanchurin}
\AuthorCitation{Gusev Ya.; Vanchurin, V.}
\address{%
$^{1}$ \quad Artificial Neural Computing, Weston, Florida, 33332, USA\\
$^{2}$ \quad Duluth Institute for Advanced Study, Duluth, Minnesota, 55804, USA}
\fi

\abstract{We apply the physics-learning duality to molecular systems by complementing the physical description of interacting particles with a dual learning description, where each particle is modeled as an agent minimizing a loss function. In the traditional physics framework, the equations of motion are derived from the Lagrangian function, while in the learning framework, the same equations emerge from learning dynamics driven by the agent loss function. The loss function depends on scalar quantities that describe invariant properties of all other agents or particles. To demonstrate this approach, we first infer the loss functions of oxygen and hydrogen directly from a dataset generated by the CP2K physics-based simulation of water molecules. We then employ the loss functions to develop a learning-based simulation of water molecules, which achieves comparable accuracy while being significantly more computationally efficient than standard physics-based simulations.}

\ifnum\mystyleflag=0
% JCAP
\maketitle  
\else
% Entropy
 
\fi

\section{Introduction}

A general molecular system consists of atoms of various types, with each atom comprising a nucleus surrounded by electrons. The full quantum state of such a system is described by the many-body wave function \( \Psi(q,t) \), whose dynamics are governed by the time-dependent Schrödinger equation \cite{Schrodinger}:
\begin{equation}
i \hbar \frac{\partial \Psi(q, t)}{\partial t} = \hat{H} \Psi(q, t).
\end{equation}
In general, solving the many-body Schrödinger equation exactly is intractable, both analytically and numerically. A common strategy to find approximate solutions is to exploit the large mass disparity between nuclei and electrons \cite{BornOppenheimer1927}.

In the Density Functional Theory (DFT) \cite{ParrYang1989, ShollSteckel2009}, electrons are represented by the electron density whose evolution depends on the positions of the nuclei and is typically obtained by solving the Kohn-Sham equations \cite{KohnSham1965}. Under the Born-Oppenheimer approximation \cite{BornOppenheimer1927} and the adiabatic theorem \cite{Born1954}, there exists a mapping from the nuclear configuration space to the ground-state electron density. That is, for any fixed positions of nuclei, the electrons are assumed to instantaneously relax to their corresponding ground state. This allows the dynamics of the molecular system to be effectively described using only classical nuclear degrees of freedom, with the influence of the electrons incorporated via the potential energy derived from the electronic ground state. However, the approximations are only valid for a limited range of physical phenomena. For more complex processes --- such as non-adiabatic transitions, charge transfer, strong field interactions, or excited-state dynamics --- the electronic degrees of freedom cannot be effectively ``integrated out'' or reduced to a static potential. In such cases, a full treatment of electron dynamics becomes essential.

While DFT methods provide high accuracy, they remain computationally expensive and are often impractical for large-scale simulations. In contrast, classical molecular dynamics methods offer greater computational efficiency but may lack the precision required for accurately modeling certain systems. Recent developments in machine learning have shown promise in bridging this gap. In particular, machine learning models based on neural networks have been successfully employed to reconstruct potential energy surfaces~\cite{Behler2007,Behler2016,Han2018,Behler2021, Neumann2024Orb,Yang2024MatterSim}. Furthermore, recent advancements in high-dimensional neural network potentials have led to their classification into four generations~\cite{Behler2021}, each incorporating new features such as long-range interactions and non-local charge transfer effects. However, despite these advancements, state-of-the-art machine learning-based frameworks have yet to simultaneously achieve the accuracy of DFT methods and the computational efficiency of classical molecular dynamics~\cite{Mo2022}.

In this article, we propose a novel framework for modeling molecular systems by applying the physics-learning duality \cite{Vanchurin2024,vanchurin2025geometric}, where the physics description of interacting nuclei is complemented by a machine learning description of interacting agents.\footnote{A complementary approach was taken in Ref. \cite{AndrejicVanchurin2023} where a machine learning view of autonomous agents was replaced with a dual physical view of autonomous particles.} The agents (or nuclei) navigate in a complex environment of other agents (or nuclei), and the same physical equations of motion emerge from learning dynamics governed by agent loss functions. These loss functions depend on scalar quantities that describe invariant properties of other agents, allowing the system to retain key symmetries at the architectural level. While the loss function could in principle be derived from first principles, in this article we infer it directly from the CP2K simulations \cite{CP2K, cp2kgit} of water molecules. We then use the learned loss function to construct a learning-based simulation of water molecules, which runs many orders of magnitude faster than conventional physics-based molecular dynamics algorithms.

The paper is organized as follows. In Sec.~\ref{Sec:PhySystem}, we describe a general molecular system in which individual particles interact through scalar quantities (or invariants). In Sec.~\ref{Sec:Duality}, we introduce the physics–learning duality, which enables modeling the system of interacting particles as a system of interacting agents. In Sec.~\ref{Sec:LossFunction}, we explain how the agent loss function can be inferred from physics-based simulations. In Sec.~\ref{sec:Numerics}, we present numerical results of learning-based simulations of water molecules and compare them with those from physics-based simulations. Finally, in Sec.~\ref{Sec:Conclusion}, we summarize the main results of the paper.

\section{Molecular systems}
\label{Sec:PhySystem}

Consider a molecular system with $N$ nuclei at positions $q_\mu^i = (q_1^i, q_2^i, q_3^i)$, where $i \in \{1, \dots, N\}$, and whose Lagrangian is given by
\begin{equation}
\label{eq:lagrangian_classic}
L(^{\mu\nu}_{ij} q_\mu^i, \dot{q}_\nu^j) = \frac{1}{2}M_i \dot{q}_\mu^i \dot{q}^{\mu,i} - V(_{ij}d^{ij})
\end{equation}
The corresponding Euler-Lagrange equation of motion is
\begin{equation}
\ddot{q}_\mu^k = - \frac{1}{M_k} \frac{\partial}{\partial q^{\mu}_{k}}  V(_{ij}d^{ij}) \label{eq:eom}
\end{equation}
with the separation distance
\begin{equation}
d^{ij} = \sqrt{(q_\mu^i - q_\mu^j)(q^{\mu,i} - q^{\mu,j})}.
\end{equation}
Note that throughout the paper we shall use a rather unconventional notation for representing the arguments $q^i_\mu$ of a function $f (\cdot)$ as 
\be
f (_i^\mu q^i_\mu)  \equiv f (q^1_1, q^1_2, q^1_3, q^2_1, q^2_2, q^2_3,  ..., q^N_1, q^N_2, q^N_3),
\ee 
which reduces to Einstein's summation convention over repeated indices only for linear functions, i.e., 
\begin{equation}
f (_i^\mu q^i_\mu) = f_i^\mu q^i_\mu = \sum_{i,\mu}  f_i^\mu q^i_\mu.
\end{equation}

If there are $A$ types of nuclei then we can define a many-to-one function ${\cal A}_i \in \{1,...,A\}$ whose preimage ${\cal A}^{-1}(a)$ is a set of all particles of type $a$. Then, for the two-body interactions, the potential energy can be rewritten as  
\begin{equation}
\label{eq:2_point_pot}
V(_{ij}d^{ij}) = \sum_{i,j} {\cal V}_{{\cal A}_i {\cal A}_j} \left(d^{ij} \right) = \sum_{a,b}\;\;\; \sum_{i \in {\cal A}^{-1}(a), j \in {\cal A}^{-1}(b)} {\cal V}_{ab} \left(d^{ij}\right).
\end{equation}
By taking the partial derivative we get
\begin{equation}
\frac{\partial V}{\partial q^{\nu}_k} = \sum_{a}  \frac{\partial}{\partial q^{\nu}_k} \sum_{i \in {\cal A}^{-1}(a)}  {\cal V}_{(a {\cal A}_k )} \left(d^{ik} \right) = \sum_{a}  \frac{\partial \varphi_{a {\cal A}_k }(^\mu q_\mu^{k})}{\partial q^{\nu}_k}
\end{equation}
where all of the complexity of the environment is encoded in the invariants
\begin{equation}
 \varphi_{a {\cal A}_k }(^\mu q_\mu^{k}) = \sum_{i \in {\cal A}^{-1}(a)}   {\cal V}_{(a {\cal A}_k )} \left(d^{i k}  \right).
\end{equation}
Note that the indices are symmetrized, i.e. ${\cal V}_{(ab)} = {\cal V}_{ab} + {\cal V}_{ba}$, and thus $ \varphi_{a b} =  \varphi_{ba}$. The corresponding equation of motion \eqref{eq:eom} is
\begin{equation}
\ddot{q}_\mu^k = - \frac{1}{M_{{\cal A}_k}} \sum_a  \sum_{i \in {\cal A}^{-1}(a)}   \frac{\partial{\cal V}_{(a {\cal A}_k )} \left(d^{i k}  \right)}{\partial q^{\mu}_{k}} \label{eq:eom1.1}
\end{equation}
or
\begin{equation}
\ddot{q}_\mu^k = - \frac{1}{M_{{\cal A}_k}} \sum_{a}  \frac{\partial \varphi_{ a{\cal A}_k}(^\nu q_\nu^{k})}{\partial q^{\mu}_{k}}.\label{eq:eom1.2}
\end{equation}
 
For many-particles interactions the environment would be encoded in the invariants
\begin{equation}
 \varphi_{a b ... {\cal A}_k }(^\mu q_\mu^{k}) = \sum_{i \in {\cal A}^{-1}(a),\, j \in {\cal A}^{-1}(b)...}   {\cal V}_{( a b ... {\cal A}_k )} \left(^n d_n^{ij ... k } \right),\label{eq:scalars}
\end{equation}
where $d_n^{ij... k}$ are the geometric invariants and the corresponding equation of motion \eqref{eq:eom}  is
\begin{equation}
\ddot{q}_\mu^k = - \frac{1}{M_{{\cal A}_k}} \sum_{a, b, ...} \sum_{i \in {\cal A}^{-1}(a),\, j \in {\cal A}^{-1}(b)...}  \frac{\partial  {\cal V}_{( a b ... {\cal A}_k )} \left(^n d_n^{ij ... k } \right)}{\partial q^{\mu}_{k}}\label{eq:eom2.1}
\end{equation}
or
\begin{equation}
\ddot{q}_\mu^k = - \frac{1}{M_{{\cal A}_k}} \sum_{a, b, ...}  \frac{\partial \varphi_{ a b ... {\cal A}_k}(^\nu q_\nu^{k})}{\partial q^{\mu}_{k}}.\label{eq:eom2.2}
\end{equation}
Evidently, the second Newton's law expressed in \eqref{eq:eom2.1} is more restrictive than in \eqref{eq:eom2.2}, because when the invariants $\varphi_{ a b ... {\cal A}_k}(^\nu q_\nu^{k})$ are not expressed through potentials ${\cal V}_{( a b \dots c)} \left(^n d_n^{ij \dots k} \right)$ \eqref{eq:scalars}, the third Newton's law does not need to apply.

Equation of motion \eqref{eq:eom2.2} can be generalized to include a non-linear function of the invariants, i.e. 
\begin{equation}
\ddot{q}_\mu^k = - \frac{1}{M_{{\cal A}_k}}  \frac{\partial}{\partial q^{\mu}_{k}}H_{\mathcal{A}_k}(_{m}^{ab...} \varphi^{m}_{ab...\mathcal{A}_k } (^\mu q^k_\mu)).\label{eq:eom_gen}
\end{equation}
For example, for two-particle interactions the non-linear generalization is given by
\begin{equation}
\ddot{q}_\mu^k = - \frac{1}{M_{{\cal A}_k}}  \frac{\partial}{\partial q^{\mu}_{k}} H_{\mathcal{A}_k}(_{m}^a \varphi^{m}_{a \mathcal{A}_k} (^\mu q^k_\mu))\label{eq:eom_final}
\end{equation}
where
\begin{equation}
\label{eq:invariant_phi_gen}
\varphi^{m}_{a \mathcal{A}_k} (^\mu q^k_\mu) = \varphi^{m}_{a \mathcal{A}_k}(\{d^{ki}, i \in \mathcal{A}^{-1}(a)\}).
\end{equation}
This is the non-linear generalization of the molecular dynamics that we shall consider in the paper. 

\section{Physics-learning duality}\label{Sec:Duality}

There are at least two complementary interpretation of the equations of motion \eqref{eq:eom_final}. One is the usual physical interpretation where there is an effective Lagrangian for all {\it particles} from which the Euler-Lagrange equations of motion are obtained. And the other one is the machine learning interpretation where the equations of motion correspond to a continuous limit of discrete steps taken by an {\it agent} in a complex environment of all other agents. As a learning system, the agent $k$ might take the following five steps:
\begin{enumerate}
    \item Scan the environment for positions of other agents $q_\mu^i$, for $i\neq k$, considered as non-trainable variables.
    \item Encode their positional information into scalar quantities  (or invariants) $\varphi^{m}_{a \mathcal{A}_k} (^\mu q^k_\mu)$.
    \item Calculate the loss function $H_{\mathcal{A}_k}(_{m}^a \varphi^{m}_{a \mathcal{A}_k} (^\mu q^k_\mu))$ for the agent's type ${\mathcal{A}_k}$.
    \item Update the force, which represents (negative of) the average gradient of the loss:
    \begin{equation}
        \tau \dot{F}_\mu^k(t) = - \frac{\partial}{\partial q^{\mu}_{k}} H_{\mathcal{A}_k}(_{m}^a \varphi^{m}_{a \mathcal{A}_k} (^\mu q^k_\mu)) - F_\mu^k(t)\label{eq:force}.
    \end{equation}
    \item Adjust trainable variables according to covariant gradient descent~\cite{guskov2025covariant}:
    \begin{equation}
        \dot{q}_\mu^k = \frac{\tau}{M_{{\cal A}_k}} F_\mu^k. \label{eq:sgd}
    \end{equation}
    where $\tau/M_{{\cal A}_k}$ is the learning rate.
\end{enumerate}
If the update is given by \eqref{eq:force} then the force is a (negative of) exponentially weighted gradient  
\begin{equation}
F_\mu^k(t) \equiv \frac{1}{\tau} \int_{-\infty}^t ds \exp\left (- \frac{t - s}{\tau} \right ) \left[ -\frac{\partial}{\partial q^{\mu}_{k}(s)} H_{\mathcal{A}_k}(_{m}^a \varphi^{m}_{a \mathcal{A}_k} (^\mu q^k_\mu(s))) \right ].\label{eq:exp}
\end{equation}
In this case, the non-local first-order equation~\eqref{eq:sgd} can be differentiated in time to yield a local second-order equation:
\begin{equation}
\tau \ddot{q}_\mu^k = -\dot{q}_\mu^k - \frac{\tau}{M_{{\cal A}_k}} \frac{\partial}{\partial q^{\mu}_{k}} H_{\mathcal{A}_k}(_{m}^a \varphi^{m}_{a \mathcal{A}_k} (^\mu q^k_\mu))\label{eq:local}
\end{equation}
which reduces to the second Newton's law~\eqref{eq:eom_final} in the limit of large $\tau$, or more precisely, when $\tau \ddot{q}_\mu^k \gg \dot{q}_\mu^k$. Note that the time-averaging is present in most modern machine learning algorithms based on the gradient descent method \cite{Cauchy1847} such as SGD \cite{robbins1951stochastic}, RMSProp \cite{hinton2012rmsprop}, Adam \cite{kingma2014adam}, AdamBelief \cite{zhuang2020adabelief}, CGD \cite{guskov2025covariant}, which allows the first-order learning system to effectively learn the higher derivatives through statistical averaging.

It is important to emphasize that the machine learning perspective is more general than the classical description of a system of particles, as a system of agents of different types need not possess a global potential energy function. On the other hand, it remains to be seen whether relevant quantum behavior can also be captured within the learning framework, potentially enabling the study of more complex molecular systems.

To summarize, in the conventional physics framework, all particles interact with one another, and these interactions are described by a Lagrangian function. In contrast, from the dual learning framework, each type of particle attempts to minimize its own loss function by adjusting its position — treated as a trainable variable — while considering the positions of other particles as non-trainable variables. Therefore to apply the physics description we must determine the right Lagrangian of all particles $L(^{\mu\nu}_{ij} q_\mu^i, \dot{q}_\nu^j) $ and to apply the learning description we must determine the loss function for each kind of particles $ H_{\mathcal{A}_k}(_{m}^a \varphi^{m}_{a \mathcal{A}_k} (^\mu q^k_\mu))$.

\section{Agent loss function}
\label{Sec:LossFunction}

In the previous section, we introduced the physics-learning duality, which highlights the two dual views of a molecular system. In the physics view, the classical dynamics is described by a Lagrangian function, while in the learning view, the dynamics is described by a loss function. Although we do not yet know the exact form of the loss function, we can attempt to infer it from experiments or numerics. The main objective of this section is to describe a covariant architecture that will allow us to learn the loss function directly from molecular simulations. Roughly speaking, we will be learning how particles learn.

Recall that the agent loss function $H_{\mathcal{A}_k}(_{m}^a \varphi^{m}_{a \mathcal{A}_k} (^\mu q^k_\mu))$ is a function of the invariants $\varphi^{m}_{a \mathcal{A}_k} (^\mu q^k_\mu)$ that are themselves functions of invariant distances \eqref{eq:invariant_phi_gen}. For example, the invariants can be defined as a sum over all particles of a given type:
\begin{equation}
\label{eq:invariant_phi_archi}
\varphi_{a}^m (^\mu q^k_\mu) = \sum_{i \in {\cal A}^{-1}(a)} w^m (d^{ki}),
\end{equation}
where $w^m(\cdot)$ are some decaying functions, meaning that the contribution to the sum is smaller if a distance between a ``source'' agent-particle $i$ and a ``sink'' agent-particle $k$ is larger. In the numerical experiments we will use three different weight functions:
\begin{itemize}
\item Exponential:
\begin{equation}
w(r) = \exp(-\alpha r),\label{eq:w_exp}
\end{equation}
\item Gaussian: 
\begin{equation}
w(r) = \exp(- \frac{r^2}{2 \sigma^2}),\label{eq:w_gauss}
\end{equation}
\item Power-law:
\begin{equation}
w(r) = \frac{1}{r^n}.\label{eq:w_power}
\end{equation}
\end{itemize}
The exponential and Gaussian functions primarily encode local environments, while power-law functions can also describe non-local interactions.

From the invariants \eqref{eq:invariant_phi_archi}, each agent \( k \) can compute the loss function \( H_{\mathcal{A}_k}(\varphi^{m}_{a}) \) and its gradient \( \frac{\partial}{\partial q^\mu_k} H_{\mathcal{A}_k}(\varphi^{m}_{a}) \) in order to determine the agent's acceleration as described by the equation of motion \eqref{eq:eom_final} which is the large-\( \tau \) limit of \eqref{eq:local}.
Of course, \textit{a priori}, we do not know the form of the agent loss function, but we can attempt to infer it directly from experiments or simulations. For example, we can train a neural network to model the loss function using training data from a physical simulations such as CP2K \cite{CP2K, cp2kgit}. 

In this paper, we shall model the agent loss function as a quadratic function of the invariants \eqref{eq:invariant_phi_archi}:
\begin{equation}
\label{eq:F_A_k_quad}
H_{\mathcal{A}_k} (_{m}^a \varphi^{m}_{a}) = {A}^{\mathcal{A}_k a}_m \varphi^{m}_{a} + \frac{1}{2} {B}_{mn}^{\mathcal{A}_k ab} \varphi^{m}_{a} \varphi^{n}_{b},
\end{equation}
where tensor ${B}_{mn}^{\mathcal{A}_k ab}$ is symmetric in terms of simultaneous permutation of index pairs $(a,m)$ and $(b,n)$, i.e. ${B}_{mn}^{\mathcal{A}_k ab} = {B}_{nm}^{\mathcal{A}_k ba}$. The corresponding equation of motion \eqref{eq:eom_final} is given by:
\begin{equation}
\label{eq:acc_F_A_k_quad}
\ddot{q}_\nu^k = - \frac{1}{M_{\mathcal{A}_k}} \left ( {A}^{\mathcal{A}_k b}_n \frac{\partial \varphi^{n}_{b} (^\mu q^k_\mu)}{\partial q^\nu_k} + {B}_{mn}^{\mathcal{A}_k ab} \varphi^{m}_{a} \frac{\partial \varphi^{n}_{b} (^\mu q^k_\mu)}{\partial q^\nu_k} \right ).
\end{equation}
Evidently, the acceleration of an agent is constructed from invariants and their gradients (covariant vectors). The tensors ${A}^{\mathcal{A}_k b}_n$ and ${B}_{mn}^{\mathcal{A}_k ab}$ can be either pre-trained using a supervised learning approach or dynamically adjusted via an unsupervised learning approach, provided that a suitable loss function is identified for the task. In this work, we adopt the former (supervised) approach.

\section{Numerical results}\label{sec:Numerics}

To demonstrate the physics-learning duality, we performed numerical experiments on a system of water molecules. First, we collected data from an \textit{ab initio} molecular dynamics simulation of water molecules using CP2K simulations \cite{CP2K, cp2kgit}. Second, we trained for a quadratic model of the agent loss functions~\eqref{eq:F_A_k_quad}, separately for oxygen (O) and hydrogen (H) atoms. Third, we used the learned agent loss functions to perform molecular dynamics simulations within the learning-based framework and compared the results to those obtained from the original physics-based simulation. For a visual demonstration, see~\cite{moleculardemo}.

\subsection{CP2K Dataset}\label{eq:dataset}

The molecular dynamics simulation of the water system was performed using CP2K software \cite{CP2K}. The prepared, ready-to-use configuration file for the simulation was taken from the official GitHub page \cite{cp2kgit}. The system consisted of 128 water molecules, modeled using the TIP5P potential under standard conditions (\(1\) bar and \(300\) K). The microcanonical ensemble was chosen, meaning that the simulation was performed with the total energy conserved. The simulation box was cubic, with lattice parameters \(a = b = c = 15.6404~\mathring{\text{A}}\), ensuring a periodic boundary condition. The simulation time step was \(0.5\) femtoseconds.

A density functional theory (DFT) method was employed for electronic structure calculations, utilizing the Pade exchange-correlation functional \cite{Wang1997Pade}. The basis sets and pseudopotentials were taken from standard CP2K libraries, with a cutoff energy of \(280\) Ry for the grid. The electronic structure convergence was achieved using orbital transformation methods with a DIIS minimizer. The results of the simulation included nuclei positions and velocities evolution with time. We used these data as a training dataset.

\begin{figure}[H]
    \centering
    \includegraphics[width=0.8\linewidth]{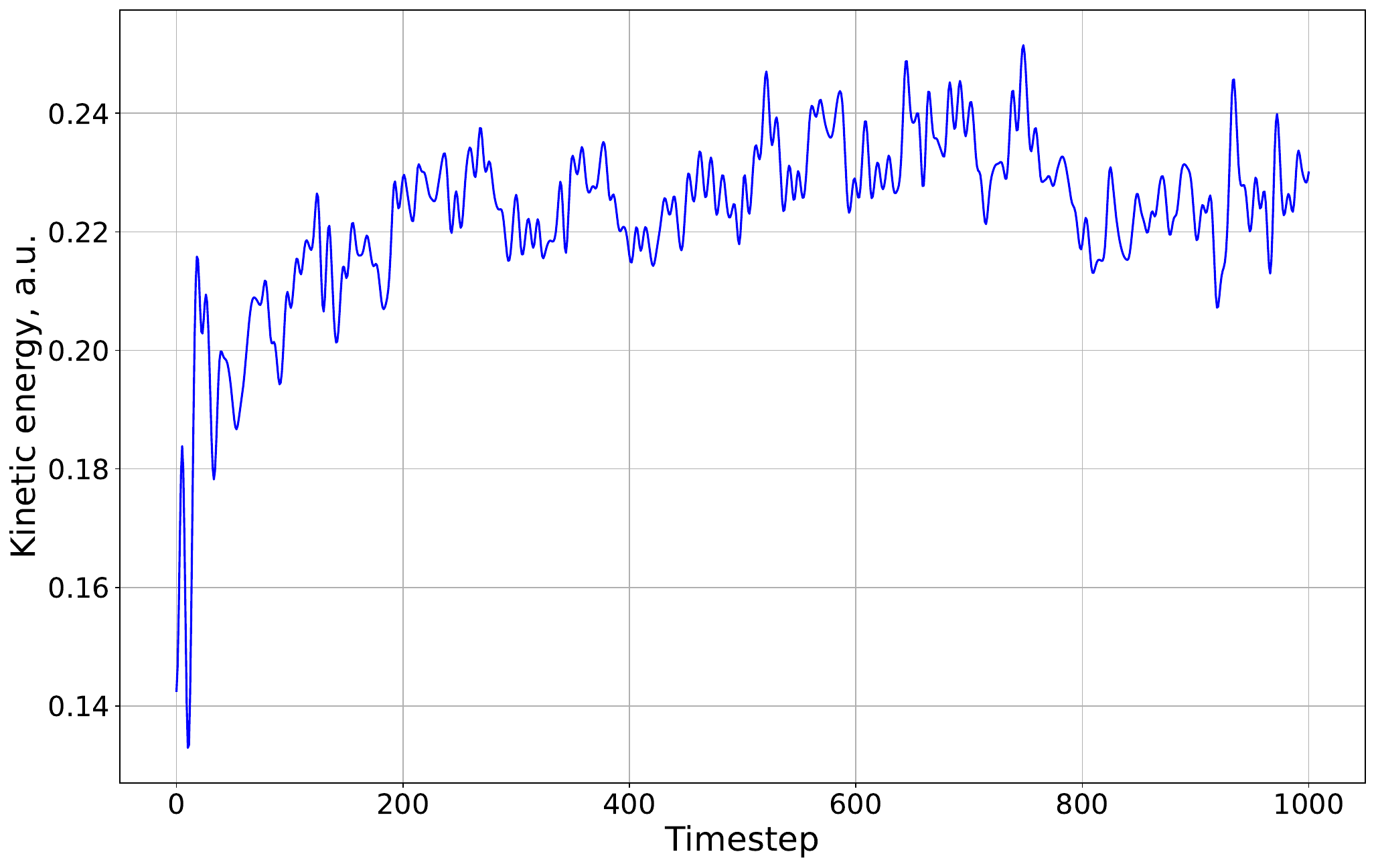}
    \caption{Kinetic energy evolution in the water simulation performed with CP2K}
    \label{fig:kin_energy_cp2k}
\end{figure}

We performed a simulation for \(1000\) time steps, but the first \(200\) time steps were excluded from the training dataset, due to the unstable kinetic energy (temperature). As we can see in Fig.~\ref{fig:kin_energy_cp2k}, the kinetic energy stabilizes only after approximately \(200\) time steps, with the respective jump in temperature from \(300\) K to approximately \(470\) K.

The data from the \(200\)-th timestep to the \(800\)-th timestep was used as the training part of the dataset, while the rest (up to the \(1000\)-th timestep) was used equally for validation and testing.

\subsection{Learning agent loss}

For constructing invariants \eqref{eq:invariant_phi_archi}, we used the following eighteen weight functions \( w^m(\cdot) \): exponential \eqref{eq:w_exp} with parameters \( \alpha = 1, \frac{1}{2}, \frac{1}{3}, \dots, \frac{1}{8} \, \mathring{\text{A}}^{-1} \), Gaussian \eqref{eq:w_gauss} with parameters \( \sigma = 1, 2, \dots, 8 \, \mathring{\text{A}} \), and power-law \eqref{eq:w_power} with powers \( n = 1, 2 \). Given the training data about the accelerations \( a^k_\nu \) of all particles in the system (see Sec. \ref{eq:dataset}), our task was to find a quadratic loss function \eqref{eq:acc_F_A_k_quad} which minimizes the mean-squared error:
\begin{eqnarray}
\label{eq:MSE_Loss_acc}
{\cal H} &=&   \left \langle \left(\ddot{q}_\nu^k - a^k_\nu  \right)^2 \right\rangle =  \left \langle \left(\frac{1}{M_{{\cal A}_k}} \frac{\partial}{\partial q^{\nu}_{k}} H_{\mathcal{A}_k}(_{m}^a \varphi^{m}_{a \mathcal{A}_k} (^\mu q^k_\mu)) + a^k_\nu  \right)^2 \right\rangle   \notag\\
&=&  \left \langle \left( \frac{{A}^{\mathcal{A}_k b}_n }{M_{{\cal A}_k}} \frac{\partial \varphi^{n}_{b} (^\mu q^k_\mu)}{\partial q^\nu_k} + \frac{{B}_{mn}^{\mathcal{A}_k ab}}{M_{{\cal A}_k}}  \varphi^{m}_{a} \frac{\partial \varphi^{n}_{b} (^\mu q^k_\mu)}{\partial q^\nu_k} + a^k_\nu  \right)^2 \right\rangle 
\end{eqnarray}
with respect to the tensors ${A}^{\mathcal{A}_k b}_n$ and ${B}_{mn}^{\mathcal{A}_k ab}$. Here, the angle brackets \( \langle \dots \rangle \) represent averaging over the entire training dataset.

\begin{figure}[h]
\begin{minipage}[h]{0.49\linewidth}
\centering
\includegraphics[width=0.99\linewidth]{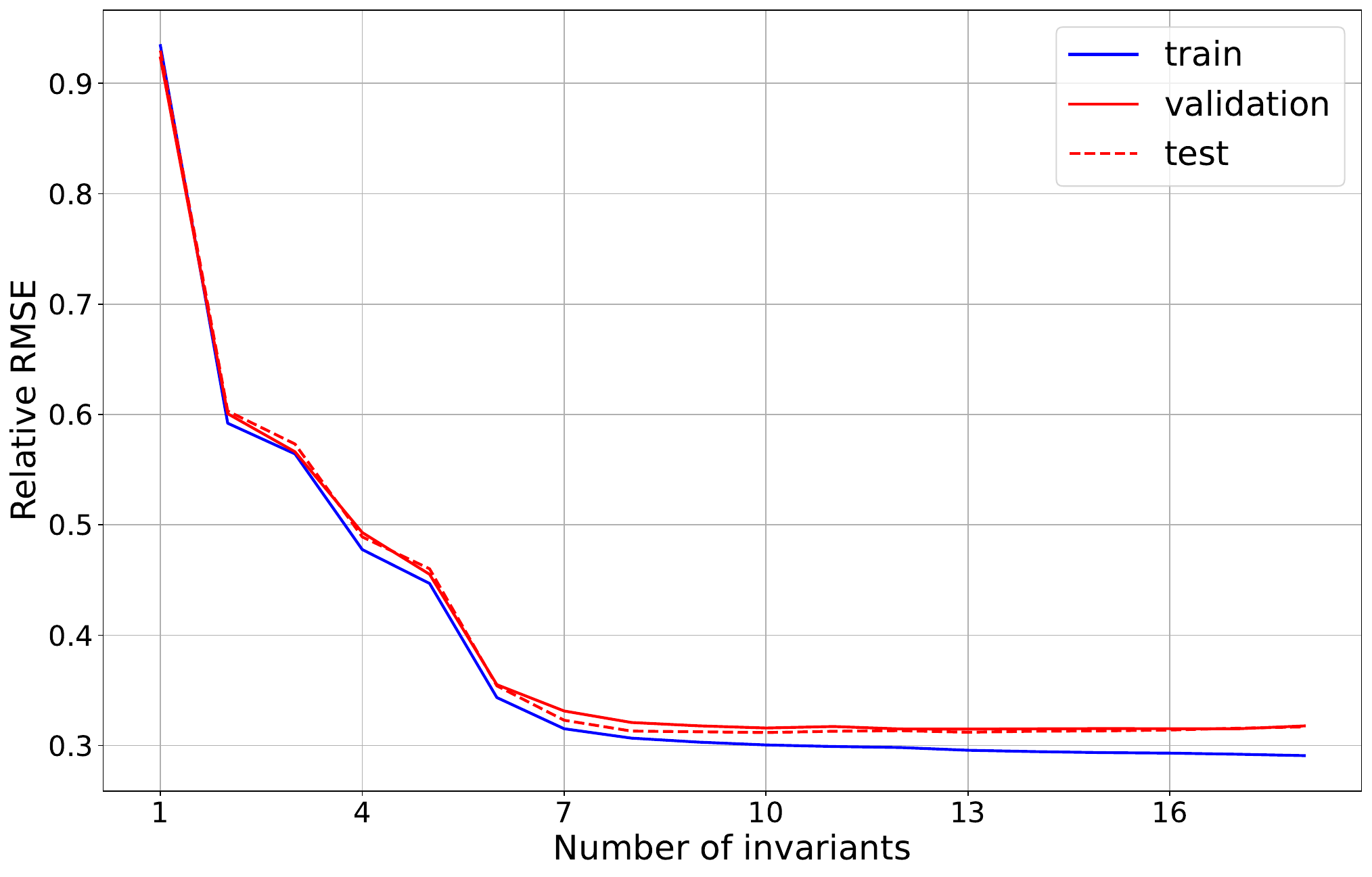} \\
Oxygen
\end{minipage}
\hfill
\begin{minipage}[h]{0.49\linewidth}
\centering
\includegraphics[width=0.99\linewidth]{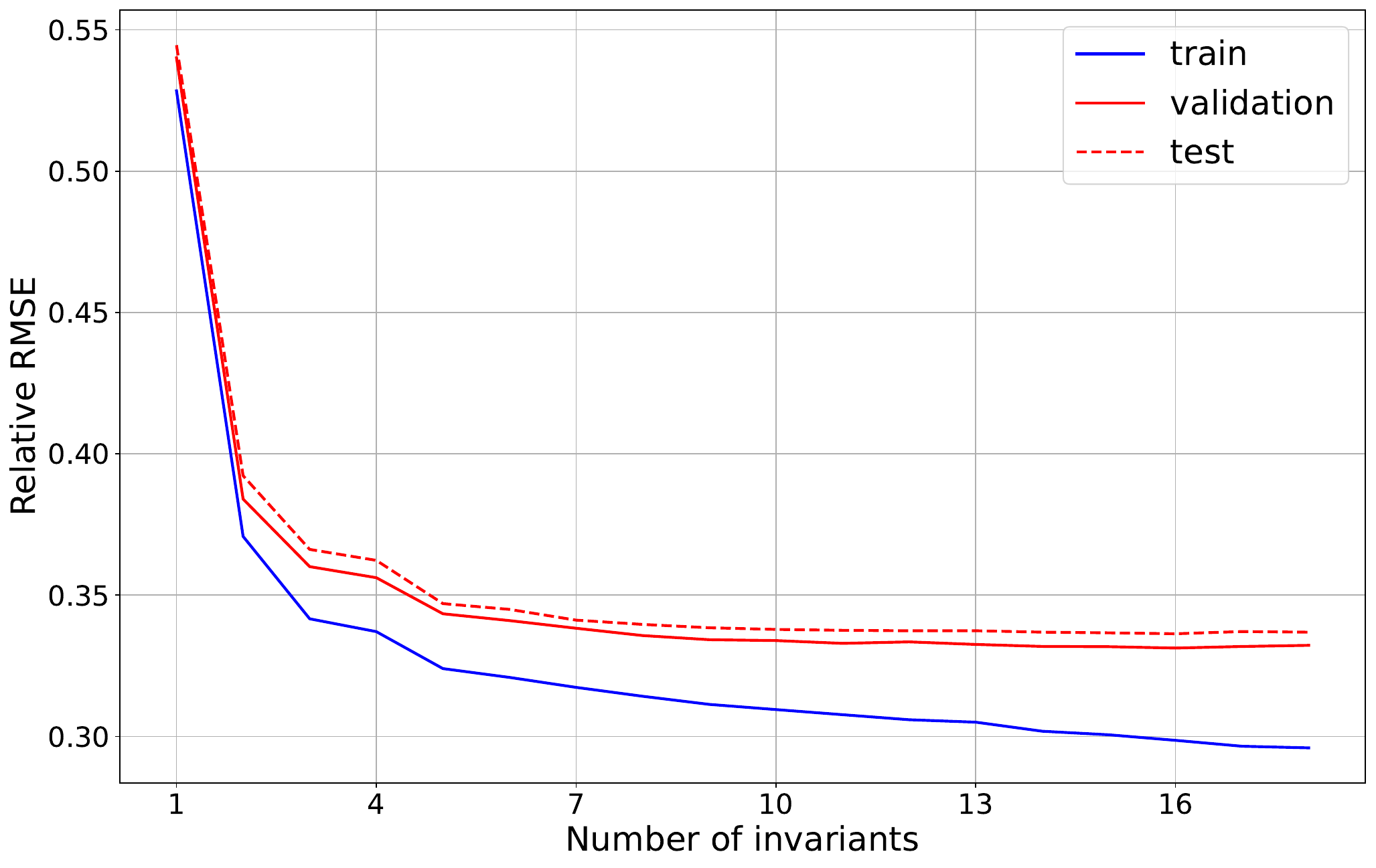} \\
Hydrogen
\end{minipage}
\caption{Relative RMSE \eqref{eq:rel_rmse} for oxygen (left) and hydrogen (right).}
\label{fig:inv_selection}
\end{figure}

To identify the relevant invariants, a process also referred to as feature selection, we begin with the complete set of available invariants. We then iteratively eliminate invariants one at a time, ensuring that the loss on the validation set remains optimal. For this analysis, we used the relative root mean-squared error (RMSE), which is a rescaled version of the loss function defined in~\eqref{eq:MSE_Loss_acc}:
\begin{equation}\label{eq:rel_rmse}
\mathcal{H}' = \sqrt{\frac{\mathcal{H}}{\left \langle (a^k_\nu)^2 \right \rangle}}.
\end{equation}
Fig.~\ref{fig:inv_selection} shows the relative RMSE as a function of the number of most relevant invariants for both oxygen and hydrogen atom types.

The seven most relevant invariants (weight functions) for hydrogen were identified as the exponential functions~\eqref{eq:w_exp} with parameters \( \alpha = \frac{1}{2}, \frac{1}{4} ~\mathring{\text{A}}^{-1} \), and the Gaussian functions with \( \sigma = 2, 3, 4, 5, 6 ~\mathring{\text{A}} \). For oxygen, the seven most relevant invariants included exponential functions with \( \alpha = \frac{1}{5}, \frac{1}{6}, \frac{1}{7} ~\mathring{\text{A}}^{-1} \), Gaussian functions with \( \sigma = 4, 6, 8 ~\mathring{\text{A}} \), and a power-law function with exponent \( n = 2 \). 

These findings indicate that hydrogen agent-particles are more sensitive to the immediate local environment, as evidenced by their preference for larger values of \( \alpha \) and smaller values of \( \sigma \). In contrast, oxygen agent-particles demonstrate a stronger responsiveness to more extended environments, reflected in smaller optimal values of \( \alpha \), larger values of \( \sigma \), and the inclusion of power-law invariants. The final training outcomes, based on these optimized sets of invariants, are presented in Table~\ref{tab:RMSE}.

\begin{table}[H]
\centering
\caption{Relative Root Mean-Squared Error \eqref{eq:rel_rmse}}
\label{tab:RMSE}
\small
\begin{tabular}{llll}
\hline
\textbf{Agents Loss Order} & \textbf{Oxygen} & \textbf{Hydrogen} \\
\hline
Linear loss \eqref{eq:F_A_k_quad} (${B}_{mn}^{\mathcal{A}_k ab} = 0$)\;\;\;\;\;\; & $0.32$ & $0.38$ \\
Quadratic loss \eqref{eq:F_A_k_quad} (${B}_{mn}^{\mathcal{A}_k ab} \neq 0$)\;\;\;\;\;\;  & $0.31$ & $0.33$ \\
\hline
\end{tabular}
\end{table}

\subsection{Molecular simulations}

Once the most relevant invariants \( \varphi^n_a \) are identified, the quadratic model of loss function~\eqref{eq:F_A_k_quad} can be employed to perform learning-based simulations of molecular dynamics. Specifically, the second-order differential equations~\eqref{eq:acc_F_A_k_quad} governing the learning dynamics can be numerically integrated using the Verlet algorithm~\cite{allen_vel_verlet}. The simulation was carried out in a canonical ensemble (NVT) setting, where the number of particles and the volume of the system were fixed, while the average temperature was maintained constant through the application of the Nosé–Hoover thermostat~\cite{nose1984,hoover1985}. In addition, we clipped the forces to prevent the agent-particles from experiencing excessive acceleration. 

\begin{figure}[H]
    \centering
    \includegraphics[width=0.8\linewidth]{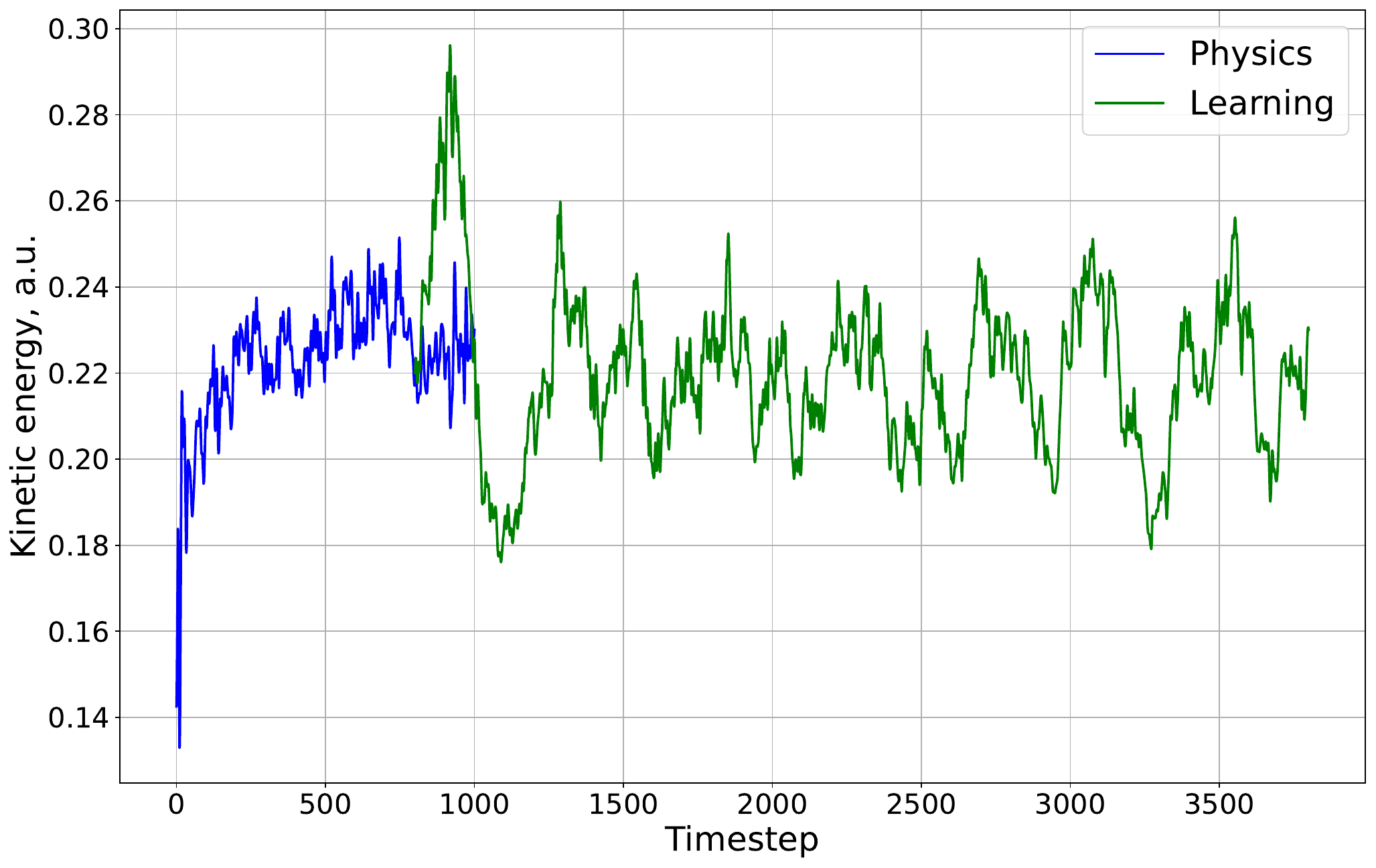}
    \caption{Time evolution of the kinetic energy in the CP2K physics-based simulation (blue) and the learning-based simulation (green).}
    \label{fig:kin_energy_cp2k_and_our}
\end{figure}

In Fig.~\ref{fig:kin_energy_cp2k_and_our}, we plot the time evolution of the kinetic energy obtained from our learning-based molecular dynamics simulation with the Nosé–Hoover thermostat, and compare it to the corresponding results from the CP2K physics-based simulation. The figure demonstrates that the learning dynamics are capable of reproducing realistic thermal behavior that is consistent with the physics-based simulation.

\begin{figure}[H]
\begin{minipage}[h]{0.49\linewidth}
\center{\includegraphics[width=0.99\linewidth]{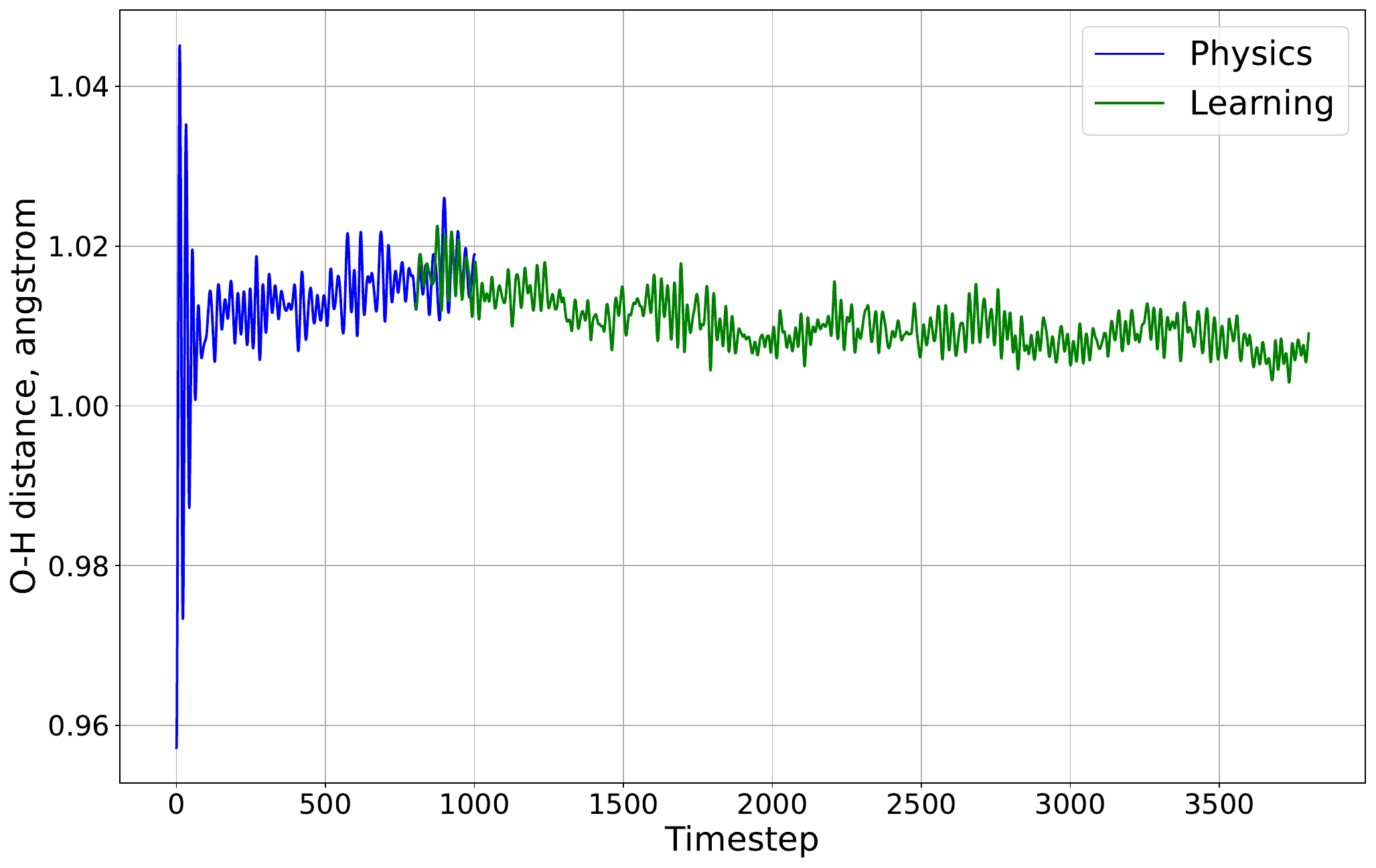} \\ Mean distance}
\end{minipage}
\hfill
\begin{minipage}[h]{0.49\linewidth}
\center{\includegraphics[width=0.99\linewidth]{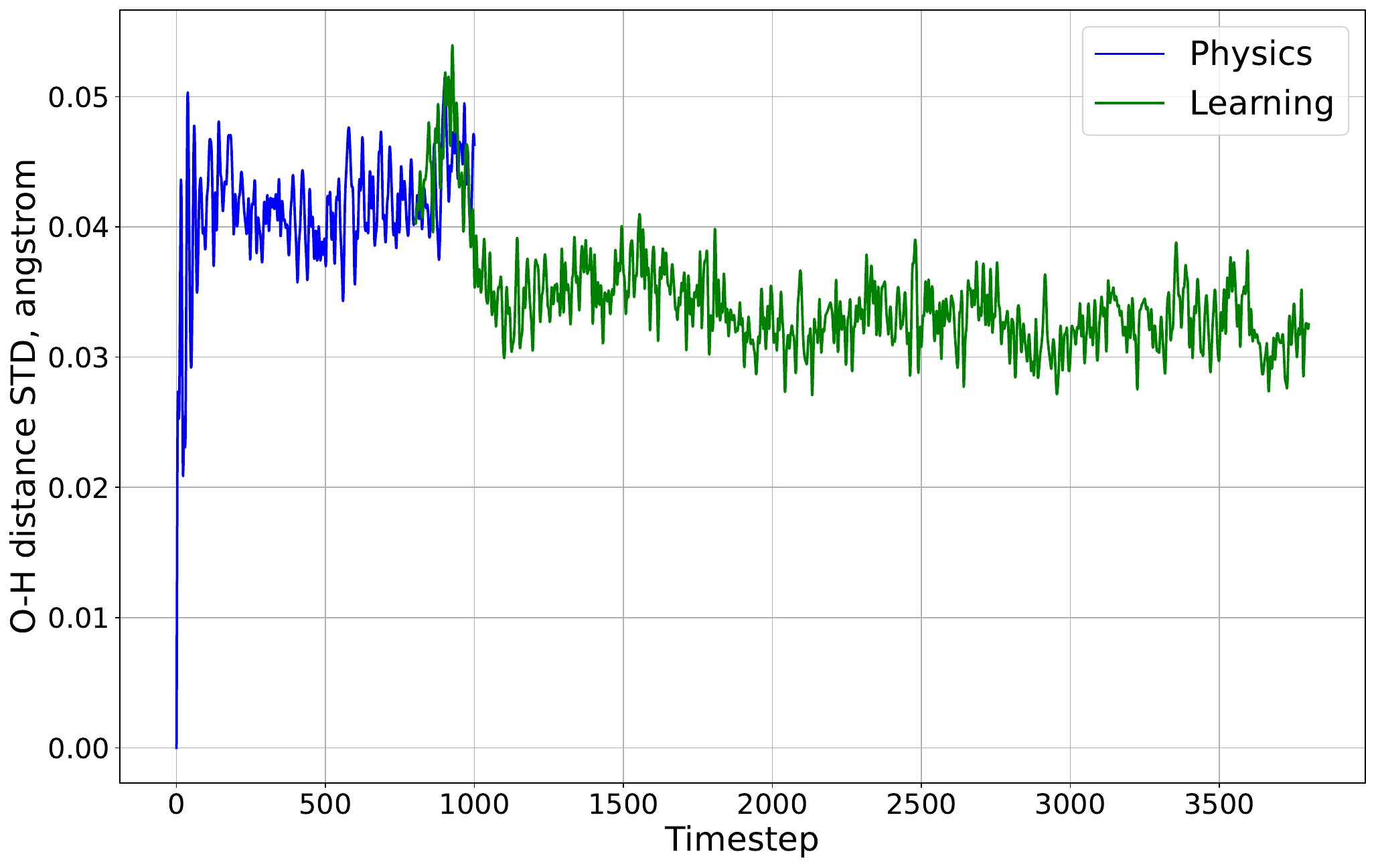} \\ Standard deviation}
\end{minipage}
\caption{Mean distance and standard deviation of the O-H bond.}
\label{fig:O_H_bond}
\end{figure}

\begin{figure}[H]
    \centering
    \includegraphics[width=0.8\linewidth]{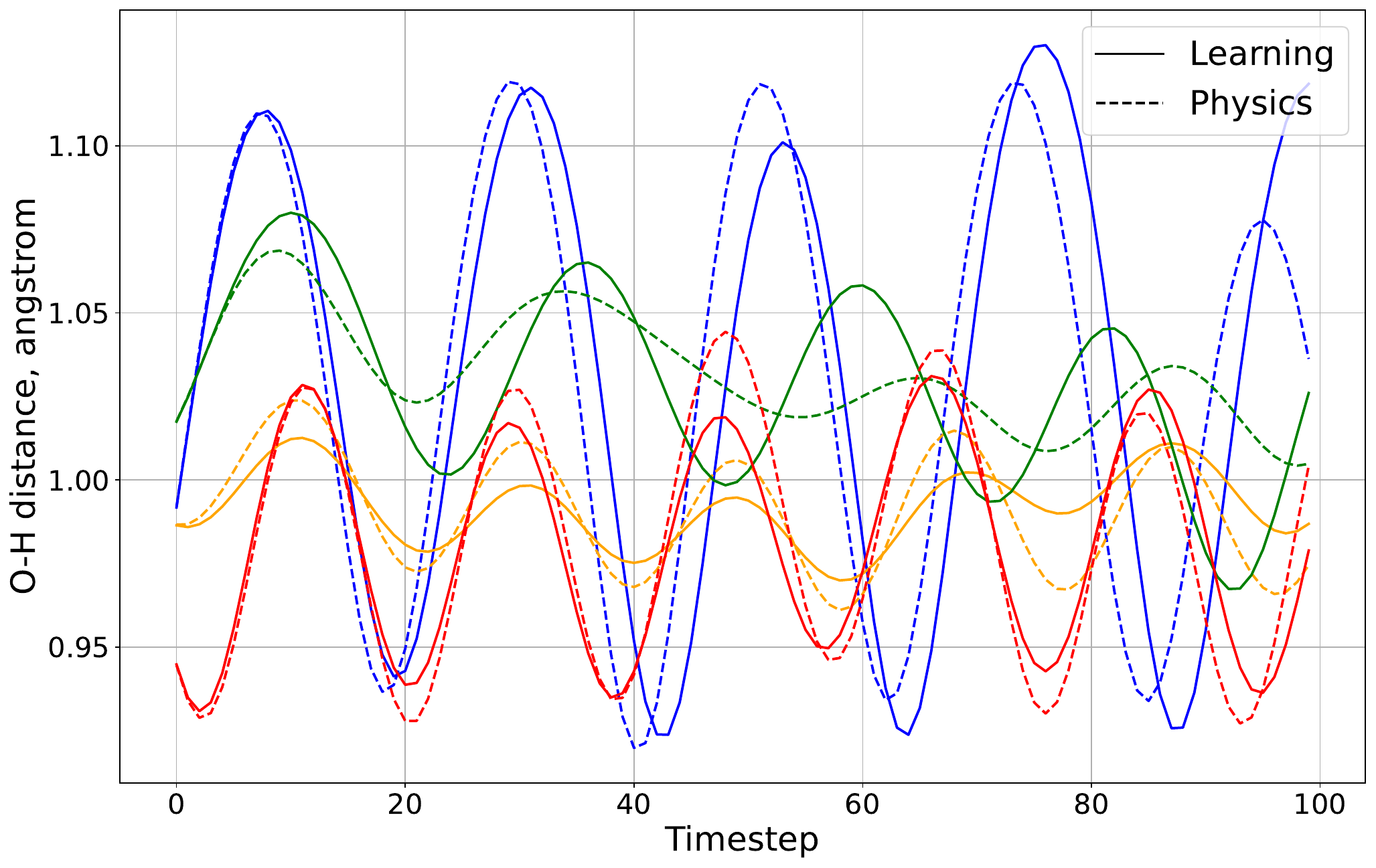}
    \caption{Time evolution of the O-H bond for four random hydrogen atoms in the CP2K physics-based simulation (dashed) and in the learning-based simulation (solid).}
    \label{fig:OH_trajs}
\end{figure}

\begin{figure}[H]
\begin{minipage}[h]{0.49\linewidth}
\center{\includegraphics[width=0.99\linewidth]{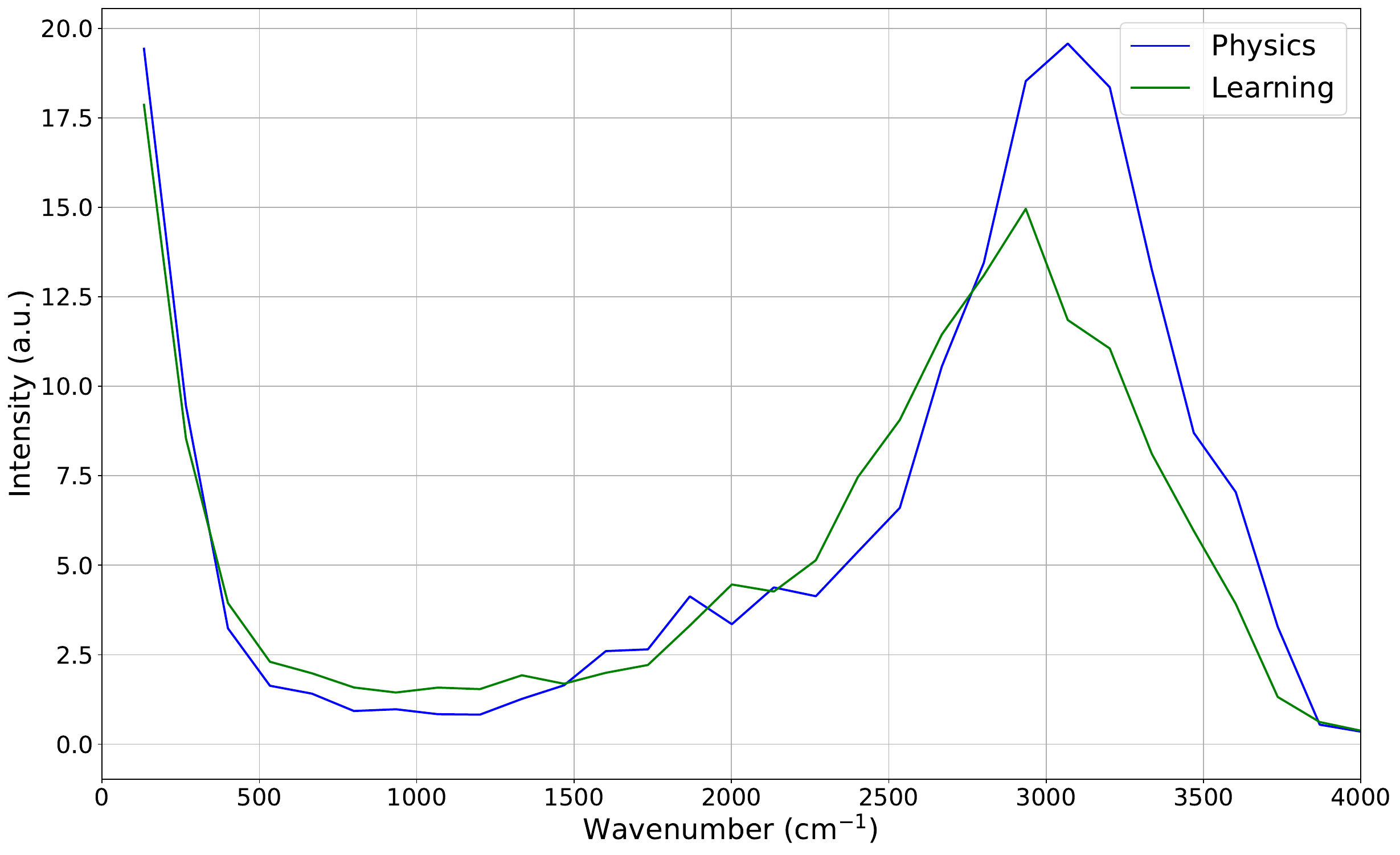} \\ Power spectrum}
\end{minipage}
\hfill
\begin{minipage}[h]{0.49\linewidth}
\center{\includegraphics[width=0.99\linewidth]{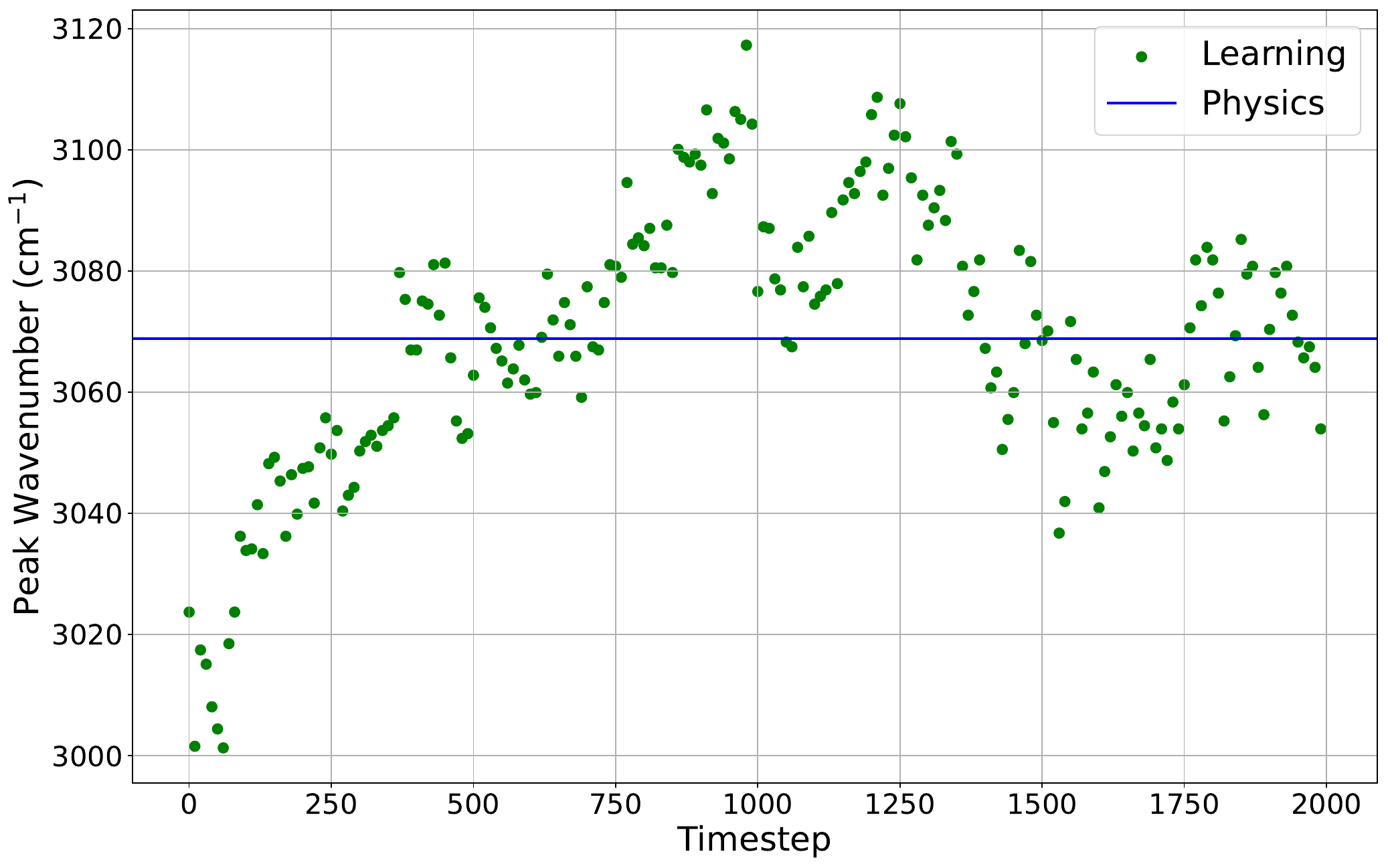} \\ Peak location}
\end{minipage}
\caption{Power spectrum from the O-H bond oscillations.}
\label{fig:OH_spectrum}
\end{figure}

\begin{figure}[H]
\begin{minipage}[h]{0.49\linewidth}
\center{\includegraphics[width=0.99\linewidth]{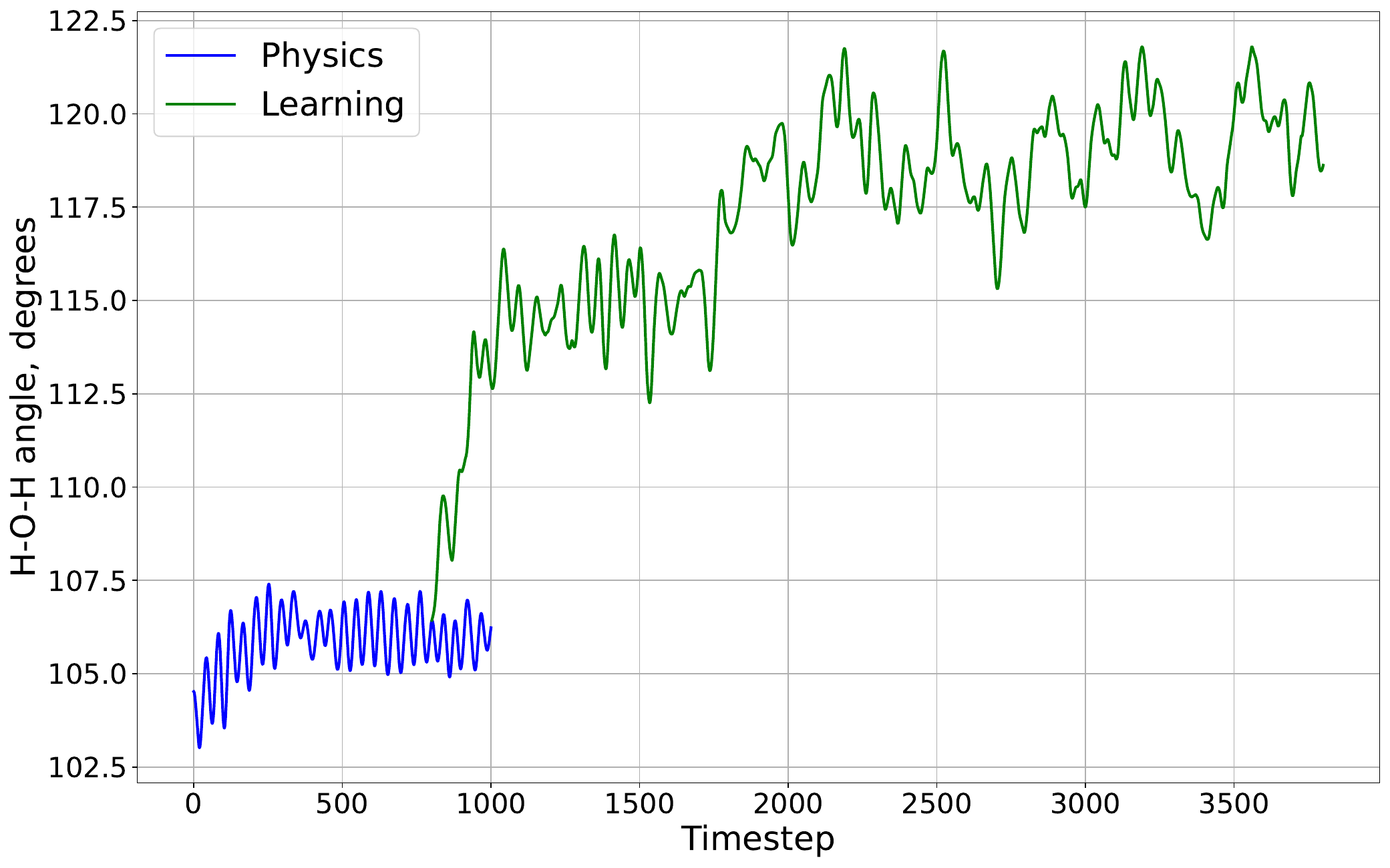} \\ Mean angle}
\end{minipage}
\hfill
\begin{minipage}[h]{0.49\linewidth}
\center{\includegraphics[width=0.99\linewidth]{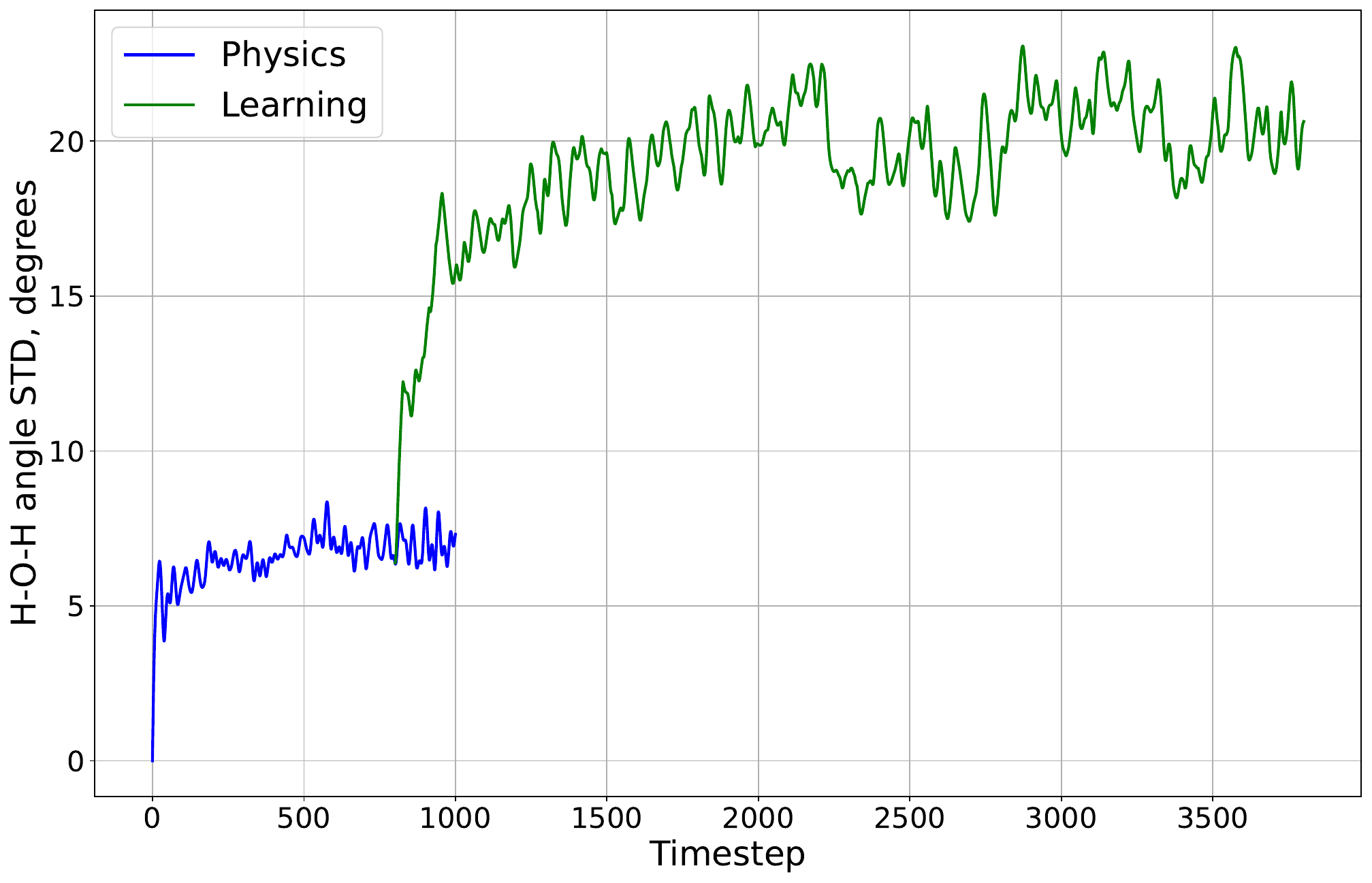} \\ Standard deviation}
\end{minipage}
\caption{Mean value and standard deviation of the H-O-H angle.}
\label{fig:HOH_angle}
\end{figure}

To further demonstrate the consistency between the physics-based and learning-based simulations, we make the following analysis:
\begin{itemize}
    \item In Fig.~\ref{fig:O_H_bond}, we plot the mean distance and the standard deviation of the O-H bond, which is comparable in both simulations.
    
    \item In Fig.~\ref{fig:OH_trajs}, we plot the time evolution of the distance between the hydrogen atom and the corresponding oxygen for four random hydrogen atoms, which are comparable in both simulations. 
    
    \item In Fig.~\ref{fig:OH_spectrum}, we plot the vibrational power spectrum and the dynamics of the spectrum's peak, which are comparable in both simulations. 
    
    \item In Fig.~\ref{fig:HOH_angle}, we plot the mean value and the standard deviation of the H-O-H angle in the water molecules, which is somewhat larger in the learning-based simulation.
\end{itemize}

The primary computational cost in the learning-based simulations arises from the evaluation of the invariants defined in Eq.~\eqref{eq:invariant_phi_archi}, which scales as $O(N^2)$ with the number of particles $N$. To mitigate this cost, one can employ fast summation techniques, such as the Particle Mesh Ewald method~\cite{Darden1993PME} or the Barnes--Hut algorithm~\cite{Barnes1986}, which reduce the complexity to $O(N \log N)$. For comparison, executing 1000 time steps of the CP2K physics-based simulation of 128 water molecules on an AMD Ryzen 9 7950X processor required approximately 10 hours, whereas the learning-based simulation completed the same number of steps in only 20 seconds.

\section{Conclusion}
\label{Sec:Conclusion}

The results of our study demonstrate that the physics-learning duality offers a promising framework for modeling molecular systems, where the dynamics of particles are effectively described as learning dynamics of interacting agents. By directly inferring the agent loss functions from CP2K simulations, we showed that the learning-based approach successfully reproduces key features of water molecule dynamics, such as bond lengths, vibrational spectra, and thermal fluctuations. Although the quadratic loss function model is relatively simple, it effectively captures the essential interactions between oxygen and hydrogen atoms, with invariants reflecting the varying sensitivities of each particle type to its local environment. Notably, the learning-based simulations achieved a substantial computational speedup over traditional physics-based simulations, underscoring the potential of this approach for large-scale simulations.

One of the key advantages of the learning-based framework is its flexibility in modeling complex interactions without the need for an explicit global potential energy function. While traditional physics-based simulations rely on predefined potentials or computationally expensive quantum mechanical calculations, the learning-based approach learns effective interactions directly from local invariants. This makes it particularly well-suited for systems where the potentials are difficult to derive or computationally prohibitive. However, the current implementation assumes pairwise interactions and employs a quadratic loss function, limiting its ability to capture the more complex many-body effects. In future work, we plan to explore higher-order loss functions and incorporate non-local invariants to further enhance accuracy.

On a more speculative note, this work reinforces the correspondence between physical and learning systems \cite{Vanchurin2024, vanchurin2025geometric}, and supports the hypothesis that the universe itself may operate as a learning system such as a neural network \cite{Vanchurin2, TTML, alexander2021autodidactic, vanchurin2021towards}. The implications of this work extend well beyond molecular simulations, offering a conceptual foundation for rethinking the dynamics of complex systems across physics \cite{Katsnelson2023EmergentScale, kukleva2024dataset}, biology \cite{Vanchurin2022MultilevelLearning, Vanchurin2022ThermodynamicsEvolution}, and beyond. This perspective suggests that learning dynamics may play a pivotal role in understanding the self-organization of complex systems across all length, time, and energy scales. We defer the exploration of these and related questions to future research.

{\it Acknowledgements.} The authors are grateful to Ekaterina Kukleva, Mikhail Katsnelson and Kirill Zatrimaylov for many stimulating discussions.

\bibliographystyle{unsrt}
\bibliography{library}

\end{document}